\documentclass[twocolumn,superscriptaddress,prl,aps,showpacs]{revtex4}
\usepackage{graphicx}
\usepackage{dcolumn}
\usepackage{bm}

\begin{document}

\title{Momentum Distribution of Near-Zero-Energy Photoelectrons in the
Strong-Field Tunneling Ionization in the Long Wavelength Limit}
\author{Q. Z. Xia}
\affiliation{National Laboratory of Science and Technology on Computational Physics,
Institute of Applied Physics and Computational Mathematics, Beijing 100088,
China \\
}
\author{D. F. Ye}
\affiliation{National Laboratory of Science and Technology on Computational Physics,
Institute of Applied Physics and Computational Mathematics, Beijing 100088,
China \\
}
\author{L. B. Fu}
\email{lbfu@iapcm.ac.cn}
\affiliation{National Laboratory of Science and Technology on Computational Physics,
Institute of Applied Physics and Computational Mathematics, Beijing 100088,
China \\
}
\author{X. Y. Han}
\affiliation{National Laboratory of Science and Technology on Computational Physics,
Institute of Applied Physics and Computational Mathematics, Beijing 100088,
China \\
}
\author{J. Liu}
\email{liu_jie@iapcm.ac.cn}
\affiliation{National Laboratory of Science and Technology on Computational Physics,
Institute of Applied Physics and Computational Mathematics, Beijing 100088,
China \\
}
\affiliation{HEDPS, Center for Applied Physics and Technology, Peking University, Beijing
100084, China\\
}

\begin{abstract}
We investigate the ionization dynamics of Argon atoms irradiated by an
ultrashort intense laser of a wavelength up to 3100 nm, addressing the
momentum distribution of the photoelectrons with near-zero-energy. We find a
surprising accumulation in the momentum distribution corresponding to meV
energy and a \textquotedblleft V"-like structure at the slightly larger
transverse momenta. Semiclassical simulations indicate the crucial role of
the Coulomb attraction between the escaping electron and the remaining ion
at extremely large distance. Tracing back classical trajectories, we find
the tunneling electrons born in a certain window of the field phase and
transverse velocity are responsible for the striking accumulation. Our
theoretical results are consistent with recent meV-resolved high-precision
measurements.
\end{abstract}

\maketitle

\emph{Introduction}---The above-threshold ionization (ATI) phenomenon of
atoms exposed to a strong field has attracted sustaining attention for
decades since it was first discovered in 1979 \cite{Agostini1979}. One of
the most pronounced features of the ATI in the long-wavelength limit, is the
high-energy photoelectron spectrum plateau extending up to 10$U_{P}$ ($%
U_{P}=I/4\omega^{2}$, denotes the ponderomotive energy, where $I$ is the
laser intensity and $\omega$ the frequency in atomic units) \cite{plateau}.
The underlying mechanism has been attributed to the tunneled electron's
multiple return and rescattering by its parent ion \cite{wbecker, LiuJ,
Corkum2011,Paulus1994,tong2007}.

Recently, some unexpected low-energy structures (LES) of ATI \cite%
{CIBLAGA2009,QuanW2009,CYWU2012} were observed in the tunneling regime,
initially at several eV and then at lower energies of less than 1eV,
triggering a new surge of attention of ATI. Although from all aspects,
including quantum and semiclassical, numerous theoretical investigations %
\cite%
{CIBLAGA2009,QuanW2009,CYWU2012,LiuC2010,YanTM2010,KastnerA2012,BurenkovI2010,CLemell2012,FCatoire2009,FHMFaisal2009}
of the underlying physics of the LES were reported, the controversies on the
surprising structure have been continuing. In the above experiments, the
photoelectrons are detected only in a small angle around the laser
polarization direction by a time-of-flight spectrometer. Because the LES is
subtle and sensitive, it is difficult to disentangle the origin of the
various findings without large solid angle measurements including momentum
information with high resolution. Recent experimental work by J. Dura \emph{%
et al} \cite{Dura_2013} steps forward in this direction. In the experiment,
with a specifically developed ultrafast mid-IR light source of 3100 nm in
the combination with a 3D reaction microscope, the strong-field dynamics is
explored in three-dimensional momentum space down to meV electron energies
with an unprecedented precision \cite{Dura_2013}. Instead of structures on
the eV level, an apparent meV electron accumulation in the ATI spectrum and
a striking momentum distribution for the near-zero-energy electron are
observed. Nevertheless, the physics
underlying the meV electron distribution is unsettled that calls for an
investigation from theoretical side urgently.

In this Letter, stimulated by the recent experiment and attempting to
resolve the controversies on LES, we theoretically investigate the
ionization dynamics of Argon atoms by intense laser fields in the deep
tunneling regime of $\gamma \ll 1$ (Keldysh parameter $\gamma =\sqrt{%
I_{p}/2U_{p}}$, where $I_{p}$ is the ionization potential), with special
emphasis on addressing the low momentum distribution of the near-zero-energy
photoelectrons. Our study is facilitated by an improved semiclassical
rescattering model that includes the Coulomb attraction and trajectory
interference and can precisely produce the momenta of the near-zero-energy
photoelectrons regardless of the Coulomb long-range tail. Our theory
accounts the surprising accumulation around near-zero momenta corresponding
to meV energies and predicts a \textquotedblleft V"-like structure at
slightly larger transverse momenta. We identify the roles of the Coulomb
attraction and trajectory interference, by tracing back the trajectories of
soft and chaotic scattering, respectively. Our work provides profound
insight into the meV low-energy ATI mechanism, and combined with the recent
high-precision experimental results can help unraveling the debate about the
LES.

\emph{Model}---In the semiclassical model, the atomic ionization consists
two essential physical processes, i.e., an electron tunnels through the
Coulomb field that has been dramatically suppressed by the laser field, and
the released electron is driven by laser field to scatter with its parent
ion \cite{LiuJ}. The tunneled electrons (released at a distance $r_{0}$ from
the ion) have initially zero longitudinal velocity and a Gaussian transverse
velocity distribution. Each trajectory is weighed by the ADK ionization rate
$\varpi (t_{0},v_{\perp }^{i})=\varpi _{0}(t_{0})\varpi _{1}(v_{\perp }^{i})$
\cite{ADK}, where $\varpi _{1}(v_{\perp }^{i})=(2\sqrt{2I_{p}}v_{\perp
}^{i}/|\varepsilon (t_{0})|)\exp [-\sqrt{2I_{p}}(v_{\perp
}^{i})^{2}/|\varepsilon (t_{0})|]$ is the distribution of initial transverse
velocity, and $\varpi _{0}(t_{0})=|\varepsilon (t_{0})|^{(1-2/\sqrt{2I_{p}}%
)}\exp [-2(2I_{p})^{3/2}/|3\varepsilon (t_{0})|]$, depending on the field
phase $\omega t_{0}$ at the instant of tunneling as well as on the
ionization potential $I_{p}$. In the post-tunneling process, the electron
evolution in the combined oscillating laser field and Coulomb field is
traced via the classical Newtonian equation $d^{2}\vec{r}/dt^{2}=-\vec{r}%
/r^{3}-\vec{\varepsilon}(t)$. The physical quantities can then be calculated
through weighted averaging over the ensemble of trajectories corresponding
to diverse initial laser phases and transverse velocities at tunneling.

We have made simulations for Ar atoms with $I_{p}=0.583$ a.u.. The laser
parameters are chosen as $\varepsilon_{0}=0.053$ a.u. and $\omega=0.0147$
a.u. ($\lambda=3100$ nm) to match the experiment \cite{Dura_2013}. Thus, the
Keldysh parameter $\gamma=0.3$. The pulse envelope is half-trapezoidal,
constant for the first six cycles and ramped off within the last six cycles.
After the laser pulse is over, the instantaneous positions and momenta of
the emitted electrons are recorded. However, the instantaneous momenta are
not equal to the asymptotic ones (i.e., $r \to \infty $) collected by a
detector due to the Coulomb long-range tail. To precisely reproduce the
near-zero momentum distribution, we need to extract the asymptotic momenta
from the instantaneous positions and momenta. The Coulomb two-body system
has two conserved vectors: the angular momentum $\vec{M}=\vec{r}\times\vec{p}
$ and the Laplace-Runge-Lenz vector $\vec{A}=\vec{p}\times\vec{M}-\vec{r}/r$%
. Using the conserved quantities and after some coordinate rotations, we can
then obtain the asymptotic momenta of the emitted electrons.
\begin{figure}[t]
\includegraphics[scale=0.33]{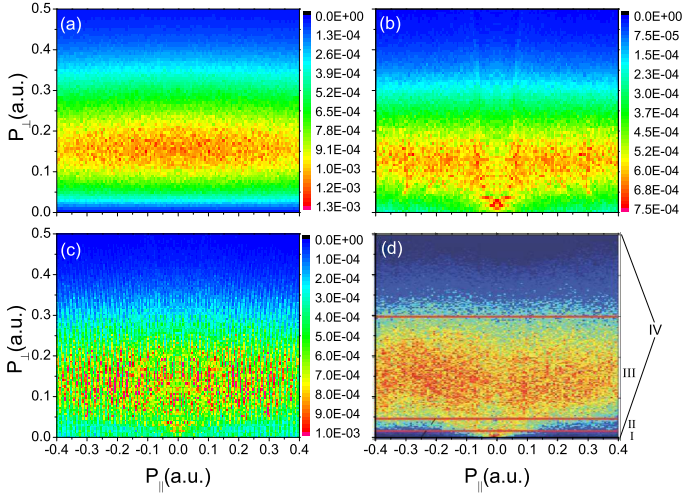}
\caption{(Color online) (a) Momentum distribution calculated from the model
without considering the Coulomb attraction in rescattering; (b) Results from
the model with considering the Coulomb attraction; (c) The results with
considering both Coulomb field and trajectory interference; (d) Momentum
spectrum from experiment cited from \protect\cite{Dura_2013} for comparison.}
\label{fig:FIG1}
\end{figure}

\begin{figure}[t]
\begin{center}
\rotatebox{0}{\resizebox *{9.0cm}{7.0cm} {\includegraphics
{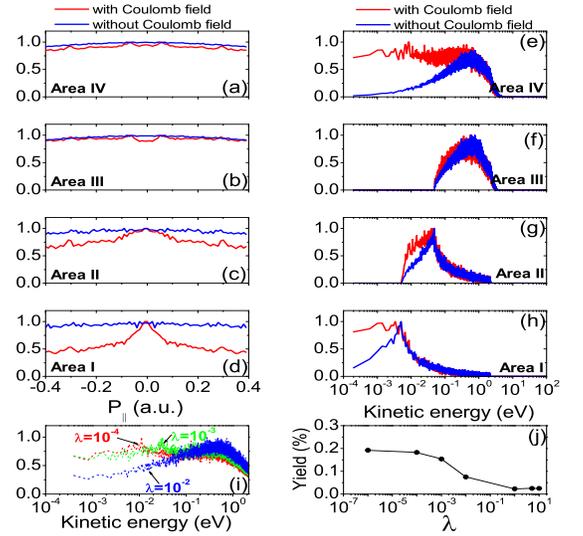}}}
\end{center}
\par
\caption{(Color online)(a)-(h)The parallel momentum distribution and the
energy distribution counted from Fig. 1(b), with respect to the four areas
within the momentum map of Fig.~\ref{fig:FIG1}(d). (a) and (e) correspond to
the area IV; (b) and (f) correspond to the area III; (c) and (g) to the area
II; and (d) and (h) to the area I. Red curves denote the results of the
semiclassical simulation with the Coulomb field, while the blue ones
represent the simulation results without the Coulomb field. (i) and (j) are
simulation results from screening potentials: (i) Low energy distribution
with respect to screening parameters; (j) meV photoelectron yields vs. the
screening parameters.}
\label{fig:FIG2}
\end{figure}

The above classical trajectory evolution, however, ignore the quantum
interference totally. To retrieve the interference effect, we assign a phase
to each trajectory. The phase $S$ is determined by the integral $%
S(t_{0},v_{\perp }^{i})=-i\int_{t_{0}}^{t_{f}}[\frac{1}{2}\vec{v}^{2}(t)-%
\frac{1}{r(t)}+I_{p}]dt$ \cite{YanTM2010}, here $\vec{v}(t)$ and $r(t)$ is
the solution of the Newton equation with the initial condition $t_{0}$, $%
v_{\perp }^{i}$ and tunneling position $r_{0}$. Then, the transition
amplitude from the initial state to the continuum state with the asymptotic
momentum $(p_{\parallel }, p_{\perp })$ can be calculated as $M(p_{\parallel
}, p_{\perp })=\sum_{t_{0},v_{\perp }^{i}}\sqrt{\varpi (t_{0},v_{\perp }^{i})%
}\exp {[S(t_{0},v_{\perp }^{i})]}$, where the summation includes all the
trajectories leading to the same final momentum $(p_{\parallel }, p_{\perp
}) $. The momentum spectra can be obtained from $|M(p_{\parallel }, p_{\perp
})|^{2}$ . Here, we only consider the electrons released within the first
cycle $(\omega t_{0}\in \lbrack 0,2\pi ])$. In our simulation, more than 5
million trajectories are calculated and the convergence of the results has
been tested by increasing the number of trajectories.

\begin{figure}[t]
\includegraphics[scale=0.38]{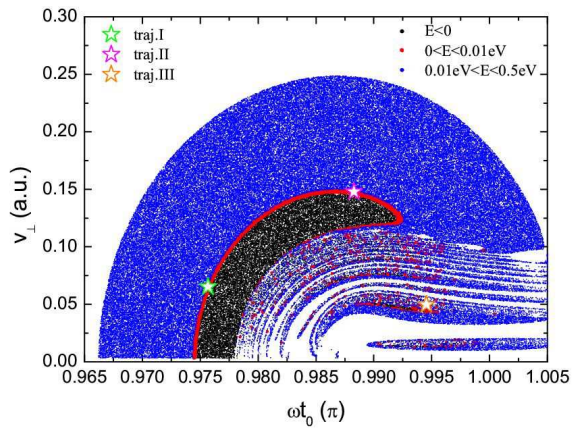}
\caption{(Color online) The dependence of final kinetic energy on tunneling
time and initial transverse velocity around peak field. The data are plotted
in different colors with respect to different electron energy ranges. The
blank areas contribute to high-energy electrons with $E>0.5$eV, which are
not in the focus of the present discussion.}
\label{fig:FIG3}
\end{figure}

\emph{Momentum Distribution of Low-Energy Electrons}--- Fig.~\ref{fig:FIG1}
(a) (b) and (c) show our model calculations on the momentum distribution
spectra of low-energy electrons in the momentum ranges $p_{\parallel} \in
(-0.4,0.4)$ and $p_{\perp} \in (0,0.5)$, indicating the prominent roles of
both the Coulomb potential and trajectory interference.

We can first see the important role of the Coulomb attraction by comparing
Fig. 1(a) and (b). In Fig. 1(a), we artificially remove the Coulomb
potential in the post-tunneling scattering. Here, the momentum spectrum
exhibits a simple \textquotedblleft sandwich"-like structure, with the dense
distribution in the central belt only reflecting the Gaussian type
distribution of initial transverse velocities. In this case, we can not see
any accumulation near zero momentum. However, in the presence of the Coulomb
potential (see Fig. 1 (b)), we can obviously observe the accumulation near
zero momentum as well as a \textquotedblleft V"-like structure at the
slightly larger transverse momenta.

\begin{figure*}[t]
\includegraphics[scale=0.5]{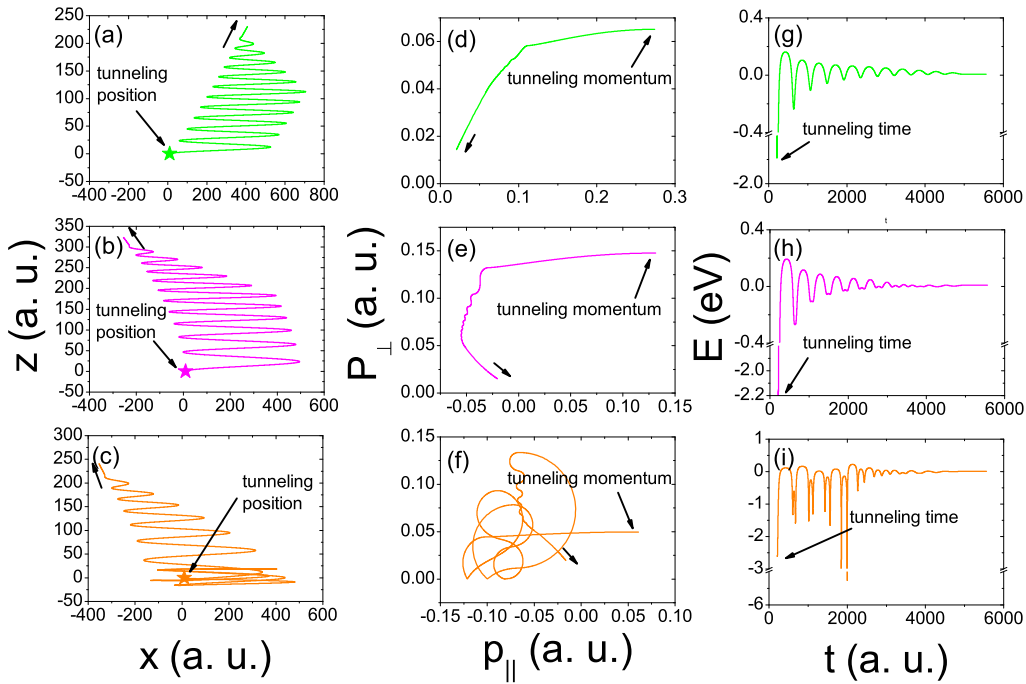}
\caption{(Color online) Three typical trajectories leading to meV energy and
their corresponding temporal evolutions of the momentum and energy. (a) (d)
and (g), the soft forward scattering trajectory; (b) (e) and (h), the soft
backward scattering; (c) (f) and (i), the chaotic scattering.}
\label{fig:FIG4}
\end{figure*}
The role of the trajectory interference becomes obvious when comparing Fig.
1 (c) and (b): The dense belt structure broadens and the accumulation near
origin fades out a little, in better agreement with the experiment. Fig.
1(c) also shows fine vertical interference contrasts.

To achieve deeper insight into the origin of the striking structure in the
momentum distribution and address its relation to the energy spectrum, we
take average the transverse momentum and generate energy distributions with
respect to distinct transverse momentum regimes, i.e., accumulation regime
(I), V-structure regime (II), and belt regime (III). Symbol IV represents
the sum over the above three regimes. In our energy statistics, the energy
interval is chosen as $0.4$ meV, consistent with the experimental
resolution. The results are shown in Fig. 2.

From Fig. 2 (f) to (h), we see the Coulomb effects increase and become very
significant at small transverse velocities. The envelope of the total energy
spectrum (e) shows a prominent hump around 0.5 eV and then extends to meV
energies or even less. The hump is also apparent in experiment
\cite{Dura_2013}. Nevertheless, it has nothing to do with the Coulomb attraction, is just
due to momentum space density compression when transformed to energy
distribution. An estimation of the hump center is given by $\partial \varpi
(t_{0},v_{\perp }^{i})/\partial t_{0}=0$ and $\partial \varpi
(t_{0},v_{\perp }^{i})/\partial v_{\perp }^{i}=0$. In the long-wavelength
limit of $\omega \rightarrow 0$, we analytically obtain an electron kinetic
energy of $\varepsilon
_{0}/4\sqrt{2I_{p}}$, here $\varepsilon _{0}$ is the maximum field strength.
Substituting the atom and laser parameters, it gives $0.33$ eV, in close
agreement with the results of the simulation. In contrast to the experiment,
below the hump, our simulation exhibits a plateau structure in energy
spectrum that spreads down the regime of $10^{-4}$ eV or less.

The meV electrons are closely related to the Coulomb attraction. In
particular, we find that the long-range Coulomb attraction between the
electron and ion plays a crucial role, in contrast to the Coulomb focusing
effect \cite{focusing} that is significant only when the electron is closer
to ion. We have replaced the Coulomb potential by a Yukawa type potential of
$-exp[-\lambda r]/r$, to screen the Coulomb long-range tail (see Fig. 2 (i)
and (j) ). We find surprisingly that, even with a very small screening
parameter of $\lambda = 0.01$, the meV electron yield decreases rapidly,
analogous to the case where the Coulomb potential is completely absent. Only
with much smaller screening parameters of $0.001$ or less, the meV electron
accumulation can recover. The above observation unambiguously indicates the
crucial role of the Coulomb attraction between the escaping electron and ion
at the extremely long distance ($\gg 100$ a.u.), and provides the strong
evidence that the highly excited Rydberg states are involved in the meV
electron dynamics. We have also calculated the meV electron yields (i. e.
energy less than 0.01 eV) with respect to the screening parameters in Fig. 2
(j), which mainly exhibits a logarithm feature. The singularity stemmed from
the Coulomb long tail also emerges in heavy ion impact ionization with
manifesting a sharp cusp-like peak at zero transverse momentum \cite{macek}.
Recently, the cutoff of Coulomb potential at the distance of a few atomic
units is found to affect the momentum spectra of the electrons with eV
energy in multiphoton regime \cite{rudenko, tong1}. Here, we find that the
tiny Coulomb tail at distance much larger than $100$ a.u. can significantly
help producing the meV photoelectrons and lead to the striking distribution
in momentum spectrum.

The striking meV electron accumulation is found to be dependent on laser wavelength.
We have extended our simulations to
shorter wavelength of $800$ nm  and find that apparent accumulation around near-zero momenta  become less
visible. This is due to the stabilization of the high-excited Rydberg states that are
deeply involved in the meV photoelectron dynamics. When field frequency
is  much larger than the Rydberg orbit frequency, the electrons that are pumped into the Rydberg states
become stabilized against
ionization \cite{partial,huang}. This effect can reduce the meV electron accumulation.
Actually, we have simply estimated the most possible energy of the electrons that are pumped by the laser field.
According to the zero origin of the energy, we can obtain a critical Keldysh parameter of
$\gamma _{c}=(\sqrt{I_{p}}/2\sqrt{2})^{1/2}$, below which
the most possible electron energy shifts to a positive value indicating the emergence of the hump structure similar to Fig. 2(e).
We therefore predict that the surprising meV electron accumulation around near-zero momenta should be universal in the
deep tunneling regime of $\gamma <\gamma _{c}$.  This is the case for the
present experiment \cite{Dura_2013}, where $\gamma =0.3$ and the critical Keldysh
parameter for Ar atom is around 0.5.

\emph{Classical Trajectory Analysis of the source of the meV electrons}---In
our semiclassical model, the emitted electron's energy is determined by the
tunneling time and initial perpendicular velocity. In Fig.~\ref{fig:FIG3},
we show the dependence of the final kinetic energy on the tunneling time and
initial transverse velocity around peak field. The meV electrons originate
from the red area that consists of a regular arc region and scattered
irregular regions below the arc. The irregular zone is self-similar and has
fractal properties \cite{LiuJ,fractal,rost,fu}. The trajectories originated
from this region might experience multiple returns to the ion and the final
energies is very sensitive to the initial conditions. The chaotic multiple
rescatterings by the Coulomb field might lead to extremely high energy
electrons that are responsible for the well-known ATI plateau structure \cite%
{LiuJ}. It can also result in extremely low-energy electrons, as shown by
the scattered red dots below the arc region.

Besides these chaotic trajectories, we find the electrons with meV energy
usually experience soft scattering and then move forward or backward \cite%
{YanTM2010,KastnerA2012}. We plot such three kinds of typical trajectories
in Fig.~\ref{fig:FIG4} (a), (b) and (c), respectively. Their initial
conditions correspond to the \textquotedblleft stars" in Fig. 3 . Temporal
evolution of the momenta and energy for each trajectory is shown in Fig.~\ref%
{fig:FIG4}(d-f) and (g-i), respectively. Since the electrons oscillate in
the laser field, here we use the compensated energy \cite{Leopold1979}
calculated via the canonical momentum instead of the kinetic energy. Fig. 4
(a) (g) and (b) (h) indicate that, some electrons originated in a certain
window of laser phase and initial transverse velocity can tunnel into
Rydberg states without ionization. They locate far away from the parent ion
and are weakly bounded by the ion's Coulomb attraction. They subsequently
experience very soft rescattering during which they can only acquire limited
field energy to be pumped into the continuum with meV energy. During the
process, the Coulomb potential attracts the electron that reduces the
electron's tunneling momentum to zero showing a kind of \textquotedblleft
friction" effect (see Fig. 4 (d) and (e)). While for the chaotic trajectory,
the electrons experience multiple scatterings with ion and "occasionally"
emit with meV energy. Our model calculation indicates both chaotic and soft
(forward or backward) recattering trajectories are the source of meV
photoelectrons and the ratio of the two kinds events (i.e., chaotic events
vs. soft rescattering events) is about $1/4$.

In summary, we have investigated the dynamics of meV ATI photoelectrons
from theoretical side for the first time. Our simulations indicate that the meV
electron generation is subtle and attributed to the extremely long-range
Coulomb tail, while it is also of universality in the deep tunneling regime. Our theoretical results account for the recent
high-precision ATI experiments and some predictions are given  that  calls for the verification from further
experiments. Since the meV energy corresponds to terahertz, the present
results can also have implications in the generation of terahertz radiation %
\cite{THZ}.

This work is supported by the National Fundamental Research Program of China
(Contact No 2011CB921503, 2013CBA01502, and 2013CB83410), the NNSF of China
(Contact Nos. 11274051, 11374040, 11078001,10933001).

\end{document}